\documentclass[prd,twocolumn,showpacs,amsmath,amssymb]{revtex4}
\usepackage{graphicx}
\usepackage{dcolumn}
\usepackage{bm}
\begin{document}

\title{Tachyon field in loop quantum cosmology: An example of traversable singularity}
\author{Li-Fang Li}
  \affiliation{Department of Physics, Beijing Normal University, Beijing 100875, China}

\author{Jian-Yang Zhu}
\thanks{Author to whom correspondence should be addressed}
  \email{zhujy@bnu.edu.cn}
  \affiliation{Department of Physics, Beijing Normal University, Beijing 100875, China}

\begin{abstract}
Loop quantum cosmology (LQC) predicts a nonsingular evolution of
the universe through a bounce in the high energy region. But LQC
has an ambiguity about the quantization scheme. Recently, the authors
in [\emph{Phys. Rev. D} \textbf{77}, 124008 (2008)] proposed a new
quantization scheme. Similar to others, this new quantization
scheme also replaces the big bang singularity with the quantum
bounce. More interestingly, it introduces a quantum
singularity, which is traversable. We investigate this novel
dynamics quantitatively with a tachyon scalar field, which gives
us a concrete example. Our result shows that our universe can
evolve through the quantum singularity regularly, which is
different from the classical big bang singularity. So this
singularity is only a week singularity.
\end{abstract}


\maketitle

\section{Introduction}
Among the candidates for a theory of quantum gravity, the
non-perturbative quantum gravity develops rapidly. In recent
years, as a non-perturbative quantum gravity scheme, Loop Quantum
Gravity (LQG) shows more and more strength on gravitational
quantization \cite{thiemann06}. LQG is rigorously constructed on
the kinematical Hilbert space. Many spatial geometrical
operators, such as the area, the volume and the length operator
have also been constructed on this kinematical Hilbert space. The
successful examples of LQG include quantized area and volume
operators \cite{rovelli95,ashtekar97,ashtekar98a,thiemann98}, a
calculation of the entropy of black holes \cite{rovelli96}, Loop
Quantum Cosmology (LQC) \cite{bojowald05}, etc. As a
successful application of LQG to cosmology, LQC has an outstanding and
interesting result---replacing the big bang spacetime singularity
of cosmology with a big bounce \cite{bojowald01}. In addition, LQC
also gives a quantum suppression of classical chaotic behavior
near singularities in Bianchi-IX models
\cite{bojowald04a,bojowald04b}. Furthermore, it has been shown
that the non-perturbative modification of the matter Hamiltonian
leads to a generic phase of inflation
\cite{bojowald02,date05,xiong1}.

Recently the authors in \cite{Mielczarek08} proposed a new
quantization scheme (we will call it $\mu_{MS}$ scheme in the following)
for LQC \footnote{The Eq.~(3) of \cite{Mielczarek08} is somewhat
confusing. In fact the new quantization scheme has nothing to do
with equation (3). The argument for the new scheme is just
expanding the standard LQC term (see Eqs.~(8)-(10)) and keeping
the terms of expansion up to the 4th order.}. In this new quantization
scheme, the classical big bang singularity is also replaced by a
quantum bounce. The most interestingly, this new scheme
introduces a novel quantum singularity. At this quantum
singularity the Hubble parameter diverges, but the universe can
evolve through it regularly, which is different from the case for the classical
big bang singularity. In order to investigate the novel quantum
singularity in this $\mu_{MS}$ scheme, we apply it to the universe
filled with a tachyon field and compared the result with the
classical dynamics and the $\bar{\mu}$ scheme which is presented
in our previous paper \cite{xiong1}.

The organization of this paper is as follows. In Sec. \ref{Sec.2}
we give out the effective framework of LQC coupled with the
tachyon field, and a brief review of the new quantization scheme
suggested in \cite {Mielczarek08}. In addition we show some
general properties of this specified effective LQC system. In Sec.
\ref{Sec.3}, we use the numerical method to investigate the
detailed dynamics of the universe filled with the tachyon field in
the $\mu_{MS}$ scheme. At the same time we compare the difference
between the $\mu_{MS}$ scheme, the $\bar{\mu}$ scheme and the
classical behavior. In Sec. \ref{Sec.4} we present some comments on
the traversable singularity. In Sec.\ref{Sec.5} we conclude and discuss our results on the $\mu_{MS}$
quantization scheme. Throughout the paper we adopt units with
$c=G=\hbar=1$.

\section{the effective framework of LQC coupled with tachyon field}

\label{Sec.2}The tachyon scalar field arises in string theory
\cite{Sen02,Sen02b}, which may provide an explanation for inflation
\cite{tachyon_inflation} at the early epochs and could contribute to
some new form of the cosmological dark energy \cite{tachyon_dark} at
late times. Moreover, the tachyon scalar field can also be used to
interpret the dark matter \cite{causse04}. In this paper we
investigate the tachyon field in the effective framework of LQC.
According to Sen \cite{Sen02}, in a spatially flat FRW cosmology the
Hamiltonian for the tachyon field can be written as
\[
H_\phi (\phi ,\Pi _\phi )=a^3\sqrt{V^2(\phi )+a^{-6}\Pi _\phi ^2}
\]
where $\Pi _\phi =\frac{a^3V\dot{\phi}}{\sqrt{1-\dot{\phi}^2}}$ is
the conjugate momentum for the tachyon field $\phi $, $V(\phi )$ is
the potential term for the tachyon field, and $a$ is the FRW scale
factor. So we have
\[
-1\leq \dot{\phi}\leq 1.
\]
Following our previous work \cite{xiong1}, we take a specific
potential for tachyon field \cite{Sen02b} in this paper,
\[
V(\phi )=V_0e^{-\alpha \phi },
\]
where $V_0$ is a positive constant and $\alpha $ is the tachyon
mass. Similar to \cite{xiong1}, we set $V_0=0.82$ and $\alpha =0.5$.
For the flat
model of universe, the phase space of LQC is spanned by coordinates $%
c=\gamma \dot{a}$ and $p=a^2$, being the only remaining degrees of
freedom after the symmetry reduction and the gauge fixing. In
terms of the connection and triad, the classical Hamiltonian
constraint is given by \cite{ashtekar03}
\begin{equation}
H_{cl}=-\frac 3{8\pi \gamma ^2}\sqrt{p}c^2+H_\phi .
\end{equation}
Considering $\bar{\mu}$ quantization scheme, the effective
Hamiltonian in LQC is given by \cite{ashtekar06b}
\begin{equation}
H_{eff,\bar{\mu}}=-\frac 3{8\pi \gamma ^2\bar{\mu}^2}\sqrt{p}%
\sin ^2\left( \bar{\mu}c\right) +H_\phi .
\end{equation}
The variable $\bar{\mu}$ corresponds to the dimensionless length of
the edge of the elementary loop and is given by
\begin{equation}
\bar{\mu}=\xi p^{-1/2},\label{mubar}
\end{equation}
where $\xi $ is a constant ($\xi >0$) and depends on the
particular scheme in the holonomy corrections. $\xi$ is given by
\begin{equation}
\xi ^2=2\sqrt{3}\pi \gamma l_p^2,
\end{equation}
where $l_p$ is the Planck length. Considering the $\mu_{MS}$
quantization scheme, the effective Hamiltonian in LQC is given by
\cite{Mielczarek08}
\begin{eqnarray}
H_{eff,\mu_{MS}}=-\frac 3{8\pi \gamma ^2}\frac{1-\sqrt{1-\frac
43\sin ^2(\bar{\mu}c)}}{\frac 23\bar{\mu}^2}\sqrt{p}+H_\phi .
\end{eqnarray}
In analogy to the $\mu_0$ scheme and the $\bar{\mu}$ scheme, we can
understand this $\mu_{MS}$ scheme as follows. We write the
effective Hamiltonian as
\begin{equation}
H_{eff,\mu_{MS}}=-\frac 3{8\pi \gamma ^2\mu_{MS}^2}\sqrt{p}%
\sin ^2\left( \mu_{MS}c\right) +H_\phi .
\end{equation}
While $\mu_{MS}$ is given by
\begin{eqnarray}
\frac {\sin ^2\left(
\mu_{MS}c\right)}{\mu_{MS}^2}=\frac{1-\sqrt{1-\frac 43\sin
^2(\bar{\mu}c)}}{\frac 23\bar{\mu}^2},
\end{eqnarray}
with $\bar{\mu}$ determined by Eq.~(\ref{mubar}). As to the
argument for the above equation we refer to \cite{Mielczarek08}.
With notation of the effective energy density $\rho _{eff}$ and the effective pressure $%
P_{eff}$, we can write the modified Friedmann equation, modified
Raychaudhuri equation and the conservation equation in the same
form
\begin{eqnarray}
&&H^2=\frac{8\pi }3\rho_{eff}, \\
&&\frac{\ddot{a}}a=\dot{H}+H^2=-\frac{4\pi }3\left( \rho
_{eff}+3P_{eff}\right) ,  \label{equation_H} \\
&&\dot{\rho}_{eff}+3H\left( \rho_{eff}+P_{eff}\right) =0.
\end{eqnarray}
In the above equations, $H\equiv \frac{\dot{a}}a$ stands for the
Hubble parameter; $\rho_{eff}$ and $P_{eff}$ are the effective
energy density and the effective pressure, respectively. For the
following three situations: the classical cosmology, the $\bar{\mu}$
scheme and the $\mu_{MS}$ scheme in LQC, we have
\begin{widetext}
\begin{eqnarray}
\rho _{eff,cl} &=&\rho _\phi =\frac V{\sqrt{1-\dot{\phi}^2}}, \\
P_{eff,cl} &=&P_\phi =-V\sqrt{1-\dot{\phi}^2}, \\
\rho _{eff,\bar{\mu}} &=&\rho _\phi \left( 1-\frac{\rho _\phi }{%
\rho _c}\right) , \\
P_{eff,\bar{\mu}} &=&P_\phi \left( 1-\frac{2\rho _\phi }{\rho _c}%
\right) -\frac{\rho _\phi ^2}{\rho _c}, \\
\rho _{eff,\mu_{MS}} &=&\rho _\phi \left( 1-\frac{\rho _\phi }{%
3\rho _c}\right) \left[ \frac 34+\frac 14\left( 1-\frac 23\frac{\rho _\phi }{%
\rho _c}\right) ^{-2}\right] , \\
P_{eff,\mu_{MS}} &=&\left[ P_\phi \left( 1-4\frac{\rho _\phi }{%
\rho _c}\right) -\frac{\rho _\phi ^2}{3\rho _c}\right] \left[ \frac 34+\frac %
14\left( 1-\frac 23\frac{\rho _\phi }{\rho _c}\right) ^{-2}\right]
\nonumber \\
&&+\frac 13\left( 1-\frac{\rho _\phi }{3\rho _c}\right) \left( 1-\frac 23%
\frac{\rho _\phi }{\rho _c}\right) ^{-3}\left( \frac{\rho _\phi ^2}{\rho _c}+%
\frac{\rho _\phi P_\phi }{\rho _c}\right) ,
\end{eqnarray}
\end{widetext}
where $\rho _c=\frac{\sqrt{3}}{16\pi ^2\gamma ^3l_p^2}$. In the
first two lines we have already used the Hamiltonian for the tachyon
field and the definitions of the energy density and the pressure
\cite{hossain05}
\begin{equation}
\rho _\phi =\frac{H_\phi}{a^3} ,\ P_\phi =-\frac{1}{3a^2}\frac{\partial H_\phi }{%
\partial a}.
\end{equation}
Correspondingly, we have the following evolution equations
\begin{widetext}
\begin{eqnarray}
classical &:&\ddot{\phi}=-\left( 1-\dot{\phi}^2\right) \frac{V^{\prime }}%
V\mp 3\dot{\phi}\left( 1-\dot{\phi ^2}\right) \left[ \frac{8\pi
}3\rho
_\phi \right] ^{1/2},  \label{ddotphi_cl} \\
\bar{\mu} &:&\ddot{\phi}=-\left( 1-\dot{\phi}^2\right) \frac{%
V^{\prime }}V\mp 3\dot{\phi}\left( 1-\dot{\phi ^2}\right) \left[
\frac{8\pi }3\rho _\phi \left( 1-\frac{\rho _\phi }{\rho
_c}\right) \right] ^{1/2},
\label{ddotphi_2} \\
\mu_{MS} &:&\ddot{\phi}=-\left( 1-\dot{\phi}^2\right) \frac{%
V^{\prime }}V  \nonumber \\
&&\mp 3\dot{\phi}\left( 1-\dot{\phi ^2}\right) \left\{ \frac{8\pi
}3\rho _\phi \left( 1-\frac{\rho _\phi }{3\rho _c}\right) \left[
\frac 34+\frac 14\left( 1-\frac 23\frac{\rho _\phi }{\rho
_c}\right) ^{-2}\right] \right\} ^{1/2}.  \label{ddotphi_4}
\end{eqnarray}
\end{widetext}
In the above equations, ``$-$'' corresponds to the expanding universe
while ``$+$'' corresponds to the contracting universe. For the
$\bar{\mu}$ scheme, the bounce happens at $\rho _\phi =\rho _c$, so
we have \cite {ashtekar06a,ashtekar06b,xiong1}
\[
\rho _\phi =\frac{V_0e^{-\alpha \phi }}{\sqrt{1-\dot{\phi}^2}}\leq
\rho _c.
\]
For the $\mu_{MS}$ scheme, the bounce happens at $\rho _\phi
=3\rho _c$, so we have \cite{Mielczarek08}
\[
\rho _\phi =\frac{V_0e^{-\alpha \phi }}{\sqrt{1-\dot{\phi}^2}}\leq
3\rho _c.
\]

\begin{figure*}[ht]
\includegraphics[totalheight=0.30\textheight,width=0.8\textwidth]{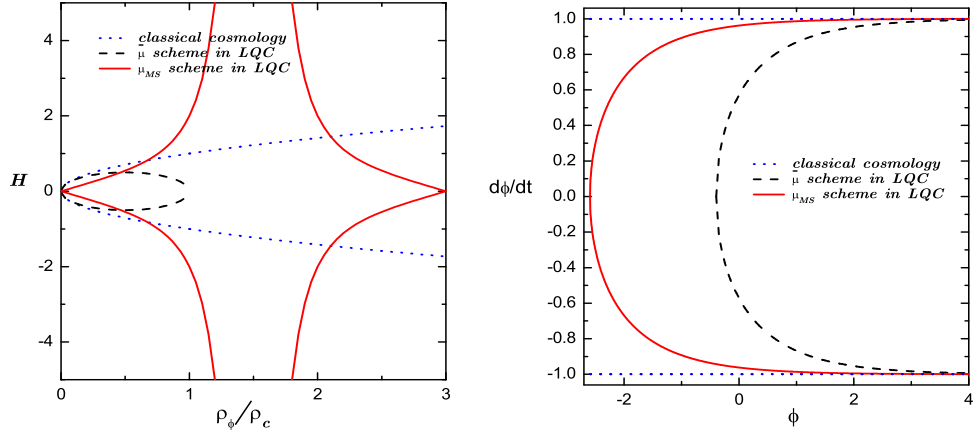}
\caption{(color online) The dotted line, dashed line and solid
line correspond to the classical, the $\bar{\mu}$ scheme and the
$\mu_{MS}$ scheme dynamics respectively. Left panel: Hubble
parameter versus
the energy density of matter. When $%
\rho _\phi \ll 1$, the quantum correction becomes negligible in
the $\bar{\mu}$ scheme, i.e., the behavior of universe goes back
to classical one. But for the $\mu_{MS}$ the quantum correction is
strong even when $\rho _\phi \ll 1$. Right panel: The phase
portrait for all admissible $\phi $ and $\dot{\phi}$ and the graph
shows the region boundary. These boundaries are nothing but the
contour lines of $\rho _\phi $ in $\dot{\phi}-\phi $ space. The
dashed line, dotted line and solid line correspond to $\infty $,
$3\rho _c$ and $\rho _c$ respectively. The admissible states for
classical system are between the two dashed lines. The admissible
states for quantum system are the right part of the corresponding
lines.} \label{fig1}
\end{figure*}

\begin{figure*}[ht]
\includegraphics[totalheight=0.40\textheight,width=0.8\textwidth]{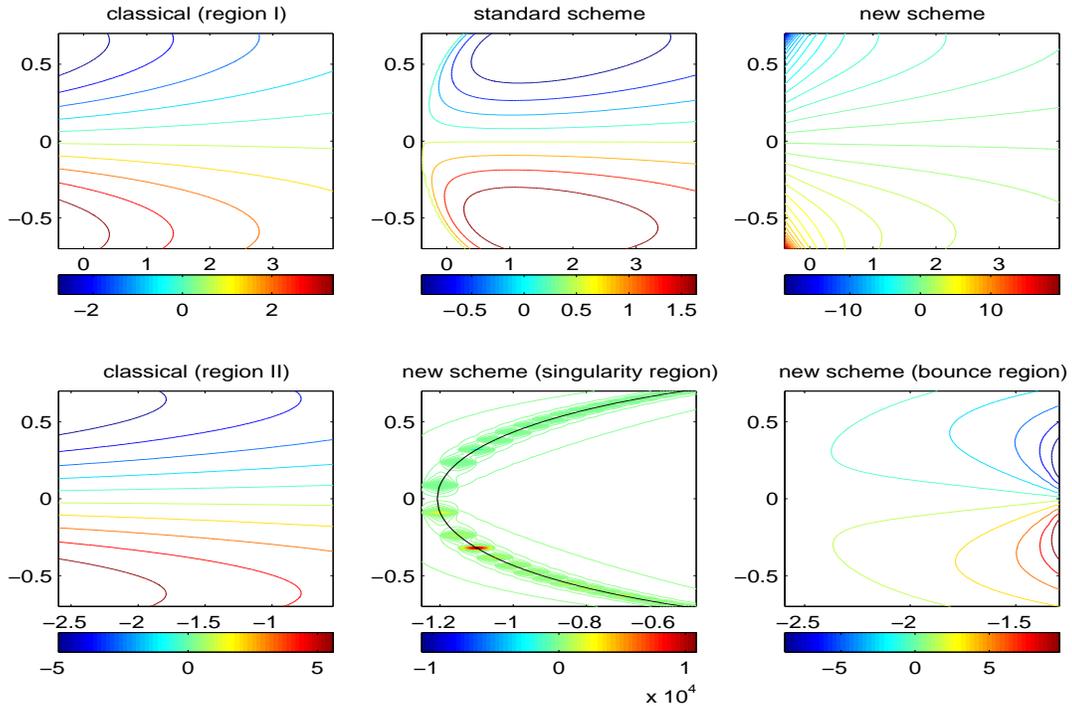}
\caption{(color online) Comparison of $\ddot{\phi}$ in the
expanding universe. The horizontal axis is $\phi $, and vertical
axis is $\dot{\phi}$. The lines are contour lines for
$\ddot{\phi}$. The upper row is comparison of three dynamical
systems near the bounce region of the $\bar{\mu}$ quantization
scheme. The lower row is comparison of the classical system and
the $\mu_{MS}$ scheme. In the three subfigures of upper row we
only show the admissible region for the $\bar{\mu}$ scheme, which
is the common region for three dynamical systems. The lines are
contour plot of $\ddot{\phi}$. The figure does not show this
singularity properly due to the numerical resolution. But the
envelopes of islands in this figure are true. At present, we are
not clear what physical phenomena these islands will result in.
Near the bounce region, the behavior of $\ddot{\phi}$ is similar
to the one of the $\bar{\mu}$ scheme in its corresponding bounce
region (compared middle subfigure in the upper row with the one of
last subfigure in the lower row).} \label{fig2}
\end{figure*}

The comparison of these features are shown in Fig. \ref{fig1} and
\ref{fig2}. In the left panel of Fig. \ref{fig1} we compare the
different behavior of the Hubble parameter versus the energy density of the matter. The upper half corresponds to the expansion
stage of the universe, and the lower half corresponds to the
contraction stage. For the $\bar{\mu}$ quantization scheme, we can
see clearly the bounce behavior at $\rho _\phi =\rho _c$. When
$\rho_\phi>\rho_c/2$ the universe meets a superinflation phase
($\dot{H}>0$). When $\rho_\phi$ becomes small, the universe
behaves as the standard picture. While for the $\mu_{MS}$ scheme,
the bounce happens at $\rho _\phi =3\rho _c$. When
$\rho_\phi>3\rho_c/2$, the universe meets a superinflation phase.
In the region where $\rho_\phi$ is small, the Hubble parameter of the
$\mu_{MS}$ scheme is smaller than the one of the standard
cosmology. In this sense the universe of the $\mu_{MS}$ scheme expands
slower than the universe of the standard cosmology.

In Fig. \ref{fig2} we compare the different behavior of
$\ddot{\phi}$ in the expanding universe (taking ``-'' in Eqs.
(\ref{ddotphi_cl})-(\ref{ddotphi_4})), in the phase space, i.e., $
\dot{\phi}$-$\phi $ space. Firstly, from the upper three subfigures,
we can see that the quantum correction in the $\bar{\mu}$ scheme of
LQC changes the amplitude of classical $\ddot{\phi}$ very small,
while in contrast, the quantum correction in the $\mu_{MS}$ scheme
changes this amplitude significantly. Besides, the quantum
correction in the $\bar{\mu}$ scheme changes obviously the line
shapes for iso-$\ddot{\phi}$, while the quantum correction in the
$\mu_{MS}$ scheme changes the line shapes negligibly. This
difference results from the different locations of the bounce
region. In fact, the quantum effect is the strongest in the bounce
region for different quantum corrections. The region shown in these
three upper subfigures is near the bounce region for the $\bar{\mu}$
scheme while some far away from the bounce region of the $\mu_{MS}$
scheme. Secondly, in the lower three subfigures, we compare the
quantum correction from $\mu_{MS}$ of LQC with the classical
behavior. We see that the behavior of $\ddot{\phi}$ near the
singularity region is changed completely. $\ddot{\phi}$ diverges
when the state approaches the singularity line (the contour line of
$\rho _\phi =\frac 32\rho _c$) except
for one point $\left( \phi ,\dot{\phi}\right) =\left( -\frac 1\alpha \ln {%
\frac{3\rho _c}{2V_0}},0\right) $ where $\ddot{\phi}=\alpha $. Yet
near the bounce region, the behavior of $\ddot{\phi}$ is similar to
the one of the $\bar{\mu}$ scheme in its corresponding bounce
region.

\section{quantitative analysis of the cosmological dynamics coupled with tachyon
field}

\label{Sec.3}
\begin{figure*}[ht]
\includegraphics[totalheight=0.30\textheight,width=0.8\textwidth]{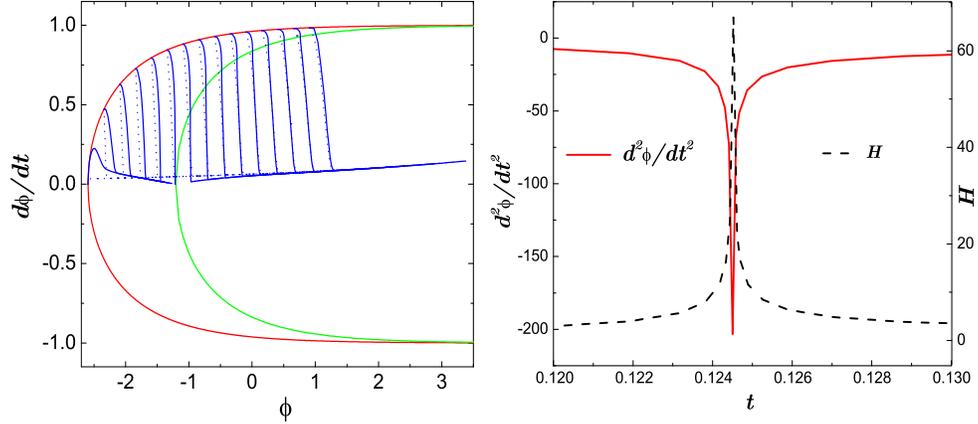}
\caption{(color online) Lept panel, comparing of the dynamical
trajectories for the classical (dotted lines) and the $\mu_{MS}$
quantization scheme (solid lines) under the expanding universe
scenario in $\dot{\phi}-\phi $ space. The left ``C'' shape line is
the boundary of admissible states for the $\mu_{MS}$ scheme,
 which is also the iso line of $\rho _\phi =3\rho _c$. The right ``C'' shape line corresponds to the singularity (the
iso line of $\rho _\phi =1.5\rho _c$) for the $\mu_{MS}$ scheme,
which is also the interface for hyper-inflationary and
deflationary states. Right panel, the concrete example of the
dynamical behavior of $\ddot{\phi}$ and the Hubble parameter $H$
near singularity for the $\mu_{MS}$ quantization scheme, which
corresponds to any line which passes through the singularity for
the $\mu_{MS}$ scheme in the left panel. The behavior of other
ones are similar on the singularity. The solid line shows behavior
of $\ddot{\phi}$, while the dotted line shows behavior of the
Hubble parameter $H=\frac{\dot{a}}a$. This subfigure is a blow-up
near singularity.} \label{fig3}
\end{figure*}

In the original paper \cite{Mielczarek08}, the
authors provided qualitative analysis of the dynamics for LQC in
the $\mu_{MS}$ scheme. Take the advantage of the specific model of
the universe coupled with the tachyon field, we can investigate
this dynamics for LQC quantitatively, and compare the difference
between the classical, the $\bar{\mu}$ scheme and the $\mu_{MS}$
scheme dynamics. We solve the equations (\ref {ddotphi_cl}),
(\ref{ddotphi_2}) and (\ref{ddotphi_4}) with the Rung-Kutta subroutine. The result is presented in Fig. \ref{fig3}.

For the $\mu_{MS}$ scheme, the difference between the classical and
the quantum behaviors is more explicit in the region between the
bouncing boundary and the singularity line. While the difference
in the region on the right-hand side of singularity is negligible. Note that
the admissible states for the $\bar{\mu}$ scheme all locate in
this region. So we can imagine that the difference between the
$\mu_{MS}$ scheme and the $\bar{\mu}$ scheme is nothing but the
difference between the classical one and the $\bar{\mu}$ one,
which is presented in Figs. 1 and 2 of our previous paper
\cite{xiong1}. Near the bounce region, the quantum behavior is
similar for both the $\bar{\mu}$ scheme and the $\mu_{MS}$ scheme.
Certainly, this behavior emerges at different places in
$\dot{\phi}$-$\phi $ space for the $\bar{\mu}$ scheme and the
$\mu_{MS}$ scheme respectively.
Near the singularity region, there is not special respects for the quantum evolution in the $%
\dot{\phi}$-$\phi $ space, except that the quantum trajectories are
much steeper than the classical ones. This steeper behavior is
resulted from the singularity behavior of the quantum dynamics which
makes $\ddot{\phi}$ much larger than the original classical ones.
From the right panel of Fig.\ref{fig3}, we can see that both the
hyper-inflationary and deflationary phases of the universe emerged
clearly. The universe expands increasingly faster before singularity
untill the acceleration becomes infinity. This stage corresponds to
the hyper-inflationary phase for the universe. After this
singularity, the universe expands more and more slowly. The behavior
comes back to the classical one \cite{sami02,guo03} quickly. This
stage is the deflationary phase for the universe.

\section{Comments on the traversable singularity}

\label{Sec.4} As a dynamical system, (\ref{ddotphi_4}) is singular
when $\rho_\phi=3\rho_c/2$, which corresponds to the right ``C"
shaped curve of the left panel of Fig. ~\ref{fig3}. That  means that
the dynamical system is only defined in two separated regions. One
is the region between the left ``C" shaped line and the right ``C"
shaped line, and the another is the region on the right-hand side of
the right ``C" shaped line. Then a question arises naturally--does
the numerical behavior traversing the separated ``C" shaped line
make sense or is it only a numerical cheating \footnote{We thank our
referee for pointing out this problem which improved our
understanding of the traversable singularity.}? If we consider these
two regions separately the dynamical system is well defined in the
sense of the Cauchy uniqueness theorem. The orbits in the phase
space $\phi$-$\dot{\phi}$ are smooth. Taking the phase space as
$R^2$, these orbits are smooth curves. So these smooth curves have
proper limit points on the separated ``C" shaped line. From the
numerical solutions shown in Fig. ~\ref{fig3}, we can see that the
different orbits have different limit points on the ``C" shaped
line. This is because the vector flow generating the trajectories
(on the $\phi$-$\dot{\phi}$ plane) has well-defined directions
(vertical) at the singularity, which means the trajectories cannot
intersect there. Then it is natural to join the two orbits in the
two separated regions with the same limit points. In this way we get
a well-defined dynamical system in the total region which lies on
the right-hand side of the left ``C" shaped line. We can expect the
numerical solution with the Rung-Kutta method will converge to this
solution (we also performed numerical integrations starting from
both sides towards the singular point and obtained the same result).
So the numerical result presented in the above sections does make
sense.

In the following, we come back to the spacetime to check the
property of this traversable singularity. Corresponding to the two
separated regions of phase space $\phi$-$\dot{\phi}$, the scale
factor $a$ can be determined by $\phi$ and $\dot{\phi}$ through
\begin{eqnarray}
a(\phi,\dot{\phi})=(\rho_\phi/H_\phi)^{1/3}\label{determinea}
\end{eqnarray}
which is a smooth function of $\phi$ and $\dot{\phi}$. Therefore,
the two spacetime regions of the universe corresponding to these two
regions is smooth. When the universe evolves to
$\rho_\phi=3\rho_c/2$, $a$ is also well defined through
(\ref{determinea}). That implies that the whole spacetime of the
universe is well defined through joining the above mentioned two
regions together by the well-defined $a$ at $\rho_\phi=3\rho_c/2$.
But since $\ddot{\phi}$ is singular there, $\dot{a}$ is also
singular there, which makes the spacetime of the universe unsmooth.
In this sense the spatial slice corresponding to this special
universe time is a traversable singularity of the spacetime.

According to \cite{ellis77,tipler77,krolak88,seifert77},
singularities can be classified into strong and weak types. A
singularity is strong if the tidal forces cause complete
destruction of the objects irrespective of their physical
characteristics, whereas a singularity is considered weak if the
tidal forces are not strong enough to forbid the passage of
objects. In this classification the singularity discussed here is
only a weak singularity. As to the cosmological singularities,
they can be classified in more details with the triplet of
variables ($a,\rho,P$)
\cite{nojiri05,cattoen05,fernandez06,singh09}: Big bang and Big
Crunch ($a=0$, $\rho$ and curvature invariants diverge at a finite
proper time); Big Rip or type I singularity ($a$, $\rho$, $P$ and
curvature invariants diverge at a finite proper time); Sudden or
type II singularity ($a$ and $\rho$ are finite while $P$
diverges); type III singularity ($a$ is finite while $\rho$ and
$P$ diverge); type IV singularity ($a$, $\rho$, $P$ and the
curvature invariants are finite while the curvature derivatives
diverge). For the singularity discussed in this paper, $a$, $\rho$
and $P$ are all finite because of the regularity of $\phi$ and
$\dot{\phi}$ at the singularity. But the Ricci curvature invariant,
\begin{eqnarray}
R=6(H^2+\frac{\ddot{a}}{a}),
\end{eqnarray}
diverges at the singularity. So the singularity here does not fall
in any type of above clarification. But it is more similar to type
IV singularity than to the other types.
\section{discussion and conclusion}

\label{Sec.5}In the classical cosmology our universe has a big bang
singularity. All physical laws break down there. LQC replaces this
singularity with a quantum bounce and the universe can evolve
through the bounce point regularly. Considering the quantum
ambiguity of quantization scheme in LQC, the authors in
\cite{Mielczarek08} proposed a new scheme. In addition to the
quantum bounce, a novel quantum singularity emerges in this new
scheme. The quantum singularity is different from the big bang
singularity and is traversable, although the Hubble parameter
diverges at this singularity. In this paper, we follow our previous
work \cite{xiong1} to investigate this novel dynamics with the
tachyon scalar field in the framework of the effective LQC. We
analyze the evolution of the tachyon field with an exponential
potential in the context of LQC, and obviously, any other choice of
potential can lead to the similar result.

In the high energy region (approaching the critical density
$\rho_c$), LQC in the new quantization scheme greatly modifies the
classical FRW cosmology and predicts a nonsingular bounce at density
$3\rho_c$, which is located at a different density region compared
with the $\bar{\mu}$ quantization scheme. In addition, besides this
quantum bounce, this new quantization scheme also introduces a
quantum singularity, which emerges at density $1.5\rho_c$. At this
quantum singularity the Hubble parameter diverges. But this
singularity is different from the classical one. The universe can
evolve through this quantum singularity regularly. Different from
the $\bar{\mu}$ scheme, the dynamics of the new quantization scheme
will deviates from the classical one even in a small energy density
region (see Fig. 1).

\acknowledgments It is a pleasure to thank our anonymous referee
for many valuable comments. The work was supported by the National
Natural Science of China (No. 10875012).

\end{document}